\documentclass[11pt]{article}
\pdfoutput=1 

\usepackage{jheppub} 

\usepackage[T1]{fontenc} 

\newcommand{\ket}[1]{|{#1}\rangle}

\newcommand{\bra}[1]{\langle{#1}|}

\newcommand{\ev}[1]{\langle{#1}\rangle}

\newcommand{\Z}{\mathbf{Z}}

\DeclareMathOperator{\area}{area}
\DeclareMathOperator{\ext}{ext}
\DeclareMathOperator{\inte}{int}
\DeclareMathOperator{\sam}{sam}
\DeclareMathOperator{\opp}{opp}

\title{General properties of holographic entanglement entropy}

\author{Matthew Headrick}

\affiliation{Martin Fisher School of Physics, Brandeis University, Waltham, Massachusetts, USA}

\abstract{The Ryu-Takayanagi formula implies many general properties of entanglement entropies in holographic theories. We review the known properties, such as continuity, strong subadditivity, and monogamy of mutual information, and fill in gaps in some of the previously-published proofs. We also add a few new properties, including: properties of the map from boundary regions to bulk regions implied by the RT formula, such as monotonicity; conditions under which subadditivity-type inequalities are saturated; and an inequality concerning reflection-symmetric states. We attempt to draw lessons from these properties about the structure of the reduced density matrix in holographic theories.}

\preprint{BRX-TH673}

\begin{document}
\maketitle
\flushbottom

\section{Introduction}

In the 1960s and 70s, the study of black holes revealed that general relativity knows thermodynamics: the geometry of spacetime encodes thermodynamic quantities like temperature and entropy, in such a way that the laws of thermodynamics become geometrical theorems.  A corollary of the statement that classical gravity is a thermodynamic theory is that quantum gravity is a statistical-mechanical theory. This idea has received its sharpest expression in the form of holographic dualities, which posit an equivalence between a gravitational theory and a field theory, where the number of fields is of order $1/(G_{\rm N}\hbar)_{\rm gravity}$, so that the classical limit of the former is the thermodynamic limit of the latter.

During the past decade, we have been learning another remarkable lesson, namely that GR also knows quantum information theory: the geometry of spacetime encodes information-theoretic quantities like entanglement entropies (EEs), in such a way that properties of quantum information like strong subadditivity become geometrical theorems. This idea has received its sharpest expression in the form of holographic EE formulas.

There are two such formulas: The Ryu-Takayanagi (RT) formula applies to static bulk states and constant-time boundary regions, giving the EE in terms of the area of a minimal hypersurface inside a bulk constant-time slice \cite{Ryu:2006bv,Ryu:2006ef}; on the other hand, the Hubeny-Rangamani-Takayanagi (HRT) formula is generally covariant with respect to both bulk and boundary diffeomorphisms, giving the EE in terms of the area of an extremal codimension-2 spacelike surface in the full Lorentzian bulk geometry \cite{Hubeny:2007xt}. In this paper we will focus on the static case.

The RT formula has been applied to a wide variety of holographic systems, leading to many insights into EEs in holographic field theories, as well as in field theories more generally. Our interest here will not be in the application of RT to any particular system, but rather in the general properties that it predicts for EEs and corresponding bulk geometrical objects. In short, whereas most work uses RT to learn about EE from holography, our goal here is to learn about holography from EE.

As we will see, there are many such properties. Some of the ones we will discuss are new. The others include ones that are explicitly the subject of previous work, ones that are implicit in the literature and/or known to experts in the field; and ones that are perhaps obvious (or at least obvious until you start thinking about them). We will attempt to provide a unified treatment, with the aim of clarifying and making explicit the various properties, the assumptions required to prove them, their logical interrelations, etc. Also, it turns out that there are gaps in the previously-published proofs of some of the previously-known properties, which we will fill as we review them.

Most of the properties we will discuss fall into two classes: (1) Ones that hold in a general quantum-mechanical system or quantum field theory, and are therefore required for the consistency of the RT formula (such as strong subadditivity); these both provide support for the formula and, presuming its correctness, give us a window into how GR encodes fundamental properties of information in spacetime geometry. (2) Ones that do not hold in a general quantum-mechanical system or field theory, but rather are special properties of holographic theories (such as monogamy of mutual information); we will attempt to give an interpretation on the field-theory side for these, again in order to learn something about holographic theories from their EEs. In a few cases (such as continuity and the reflection inequality), it is not presently known which category the property falls into; here, as in so many other instances, the power of holography has allowed us to go farther than we can for other field theories.

Given the restrictions imposed by the static EE formula, the reader might reasonably wonder why we are focusing on it, rather than the more generally applicable covariant formula. There are at least three reasons:
\begin{itemize}
\item The RT formula is far easier to work with than the HRT formula, both for calculating and for proving theorems. It therefore makes a good warm-up before tackling the much harder covariant case.
\item At the moment, the evidence in favor of the RT formula is significantly stronger than for HRT;\footnote{The evidence in favor of the RT formula includes: the fact that it satisfies a large number of required properties, as discussed in this paper; agreement with first-principles calculations of EEs in specific cases (among others, \cite{Ryu:2006bv,Ryu:2006ef,Headrick:2010zt,Casini:2011kv,Hartman:2013mia,Faulkner:2013yia}) and of the general structure of its UV divergences (among others, \cite{Solodukhin:2008dh}); and an argument relating it to Euclidean quantum gravity \cite{Lewkowycz:2013nqa}. The evidence in favor of the HRT formula includes evidence that it obeys the strong subadditivity property \cite{Allais:2011ys,Callan:2012ip,Wall:2012uf} and agreement with first-principles calculation in a much smaller number of cases (among others, \cite{Hubeny:2007xt}). The fact that both formulas have been applied in a large and diverse set of situations, apparently always giving physically reasonable results, should also be counted as evidence in their favor.} it is possible that the the former is correct while the latter is not (or at least needs to be amended or qualified in some way).
\item It is not obvious that RT and HRT always agree in cases where both can be applied. If this is not true then either one of them is wrong or they are calculating different quantities (e.g.\ the EE with respect to different states of the full system).\footnote{Examples where the two formulas apparently give different answers include bag-of-gold spacetimes \cite{Hubeny:2013gta} and certain geons \cite{Ross}.} This issue remains to be well understood.
\end{itemize}
An interesting question is whether, for each property of the static formula, an analogous property holds in the covariant case. Answering this question is important both in order to test the latter formula and (if it is correct) to learn which special properties of holographic theories extend to the time-dependent case. As we will see, the proofs in the static case are quite simple and rely only on very basic properties of minimal surfaces in Euclidean spaces (we will appeal to the Einstein equation only once, in the proof of property \ref{r(dotSigma)}). On the other hand, the analogous statements in the covariant case are novel and highly non-trivial GR conjectures. Indeed, even the very existence of an appropriate extremal surface, as required by the HRT formula, is a non-trivial conjecture, whereas in the static case the existence of a minimal surface is more or less obvious, at least at a physicist's level of rigor (as we will briefly discuss in subsection \ref{RTstatement}). Significant progress has recently been made on several of these conjectures (see \cite{Czech:2012bh,Hubeny:2012wa,Hubeny:2013gta,Hubeny:2013gba} and especially \cite{Wall:2012uf}). It is hoped that the present systematic presentation of the properties of RT might help to further such investigations.

Most of the proofs in this paper rely only on the positivity and extensivity of the area functional, together with some elementary topology. As mentioned above, we will appeal to the Einstein equation only once. In other words, the properties we discuss are in some sense kinematical, rather than dynamical. This suggests that the information-theoretic structure of spacetime implied by the RT formula applies more generally than to Einstein gravity. We leave the investigation of this issue to future work.

\subsection{Outline}

We will assume that the reader has some familiarity with the RT formula and how it is typically applied.\footnote{See \cite{Nishioka:2009un} for an overview.} Nonetheless, in section \ref{sec:setup}, we will state the formula, after explaining the general set-up and defining our notation. As emphasized there, we will strive to be as general as possible, if necessary sacrificing mathematical rigor. Therefore, our precise assumptions (concerning properties of the bulk spacetime, the region whose EE is being calculated, etc.) will vary from property to property. We will attempt to spell out the assumptions in each case, and sometimes will give counterexamples to illustrate their necessity.

We begin listing the properties in section \ref{sec:continuity}, where we will observe that the EE changes continuously under continuous variations of the region, even when bulk minimal surface whose area gives the EE jumps discontinuously. We will discuss what this phenomenon tells us about the structure of the corresponding reduced density matrices.

Throughout this paper we will emphasize an aspect of the RT formula that we believe has been generally underappreciated, namely that it associates to every spatial region $A$ of the field theory a spatial region $r(A)$ of the bulk, in a natural and canonical way. This map, which plays a role in virtually every property we will discuss, has many interesting properties in itself, which we will list in section \ref{sec:mapr}. At the end of that section we will discuss a possible interpretation of the map, namely that $r(A)$ is the holographic description of the reduced density matrix $\rho_A$ whose von Neumann entropy is the EE $S(A)$.

In section \ref{sec:inequalities} we will review inequalities derivable from the RT formula, such as subadditivity, strong subadditivity, and monogamy of mutual information. Along the way, we will fill in gaps in the previously published proofs of strong subadditivity and monogamy. We will also give a new inequality, motivated by a conjecture of Casini \cite{Casini:2010nn}, which applies when the bulk has a $\Z_2$ reflection symmetry. A very interesting feature of holographic theories is that these inequalities can be saturated, at leading order in $G_{\rm N}$, quite generically. For several of them, we will give necessary and sufficient conditions for saturation, and try to draw lessons from this phenomenon about how the field-theory degrees of freedom are organized.

\section{Background}\label{sec:setup}

\subsection{Set-up}\label{notation}

We consider a fixed, static, asymptotically anti-de Sitter spacetime $M$, which we take to be the holographic description of a static state $\rho$ of a field theory living on the conformal boundary $\dot M$.\footnote{More complicated asymptotics, such as Lifshitz geometries, should also be acceptable. This will have no effect on what follows, as long as there is a well-defined bulk Einstein-frame metric.} We work in a limit where the bulk physics is described by classical Einstein gravity. Let $\Sigma$ be a constant-time slice, and denote by $\dot\Sigma$ its conformal boundary, which is a constant-time slice of $\dot M$. We assume that $M$ (and hence $\Sigma$) is connected.

Actually, we need to introduce a regulator in order to make the areas of bulk surfaces that reach $\dot M$ finite. We will not be specific about this regulator, except that it should alter the metric on $M$ near $\dot M$ to make $\dot M$ a finite distance from points in the interior of $M$; nonetheless, the metric near $\dot M$, as well as the induced metric on $\dot M$, should in some sense be ``large''. Hence $M$ is no longer strictly speaking asymptotically AdS and $\dot M$ is no longer strictly speaking its conformal boundary. From the field-theory viewpoint, this is an ultraviolet regulator. Since we will not discuss the dependence of EEs on the regulator, we will simply leave it fixed.

The state $\rho$ may be pure or mixed (typically thermal). Indeed, if $M$ is bounded by a Killing horizon, then $\Sigma$ is bounded by its bifurcation surface $H$, and the entropy of $\rho$ is given by its area
\begin{equation}
S_{\rm tot} = \area(H)\,.
\end{equation}
(Throughout this paper, we set $4G_{\rm N}=1$, and all areas are calculated with respect to the induced Einstein-frame metric on $\Sigma$.) $H$ may be at a finite or infinite distance from points in the interior of $\Sigma$. It may also intersect $\dot\Sigma$, in which case $\dot M$ is itself bounded by a Killing horizon (as in \cite{Emparan:2006ni}). To avoid having to treat the case with no horizon separately, in that case we simply set $H=\emptyset$.

If there is a Killing horizon, then $M$ may be a subset of a larger spacetime that extends to the other side it. (The larger spaceime may either be non-static, e.g.\ if $M$ is one external region of a maximally extended AdS-Schwarzschild black hole, or static with respect to a different Killing vector, e.g.\ if $M$ is the Poincar\'e patch inside global AdS.) For the most part we will simply ignore the larger spacetime, and assume that $M$ gives a complete description of $\rho$, at least in the classical limit.

We should also consider the possibility that, in addition to horizons, $\Sigma$ is bounded by walls, where $g_{tt}$ does not vanish and the spacetime actually ends, i.e.\ there is nothing on the other side. Such walls occur naturally in many contexts; examples include confining walls, orbifold and orientifold fixed planes in string theory, and the surface $Q$ in the AdS/BCFT duality \cite{Takayanagi:2011zk}. The important difference between horizons and walls for our purposes is that the former carry intrinsic entropy (of order $1/G_{\rm N}$), while the latter don't. Walls may intersect $\dot\Sigma$ and $H$.

Presumably walls must obey some general physical constraints. Indeed, by cutting up a spacetime in arbitrary ways it is easy to produce (presumably pathological) examples that violate many of the properties we describe below. Rather than attempt to state a set of precise and general conditions on walls, we will simply point out potential pathological behaviors along the way, and where possible give assumptions that can be used to rule them out. We take the same attitude toward singularities that may occur in the interior of $\Sigma$ (branes, orbifold singularities, etc.).

As a slight generalization of the case considered here, it may be reasonable also to apply the RT formula to cases where $\Sigma$ consists of constant-time slices $\Sigma_1,\Sigma_2$ of two static regions $M_1,M_2$ joined along a common bifurcate horizon (Einstein-Rosen bridge), such as the two exterior regions of a maximally extended eternal black hole spacetime. (See \cite{Morrison:2012iz} for an example of such an application.) Although we will not consider such situations explicitly in this paper, it is fairly straightforward to see that all of the properties we describe would continue to apply.  (Note that we would not include the common horizon in $H$, since it does not bound $M$.) It can also be shown (using Properties \ref{r(dotSigma)} and \ref{monotonicity}) that such a generalization is consistent, in the sense that, for a region $A\subseteq\dot\Sigma_1$, one gets the same result for $S(A)$ whether one works in $\Sigma$ or in $\Sigma_1$.

\subsection{Regions and boundaries}\label{regions}

\begin{figure}[tbp]
\centering
\includegraphics[width=.45\textwidth]{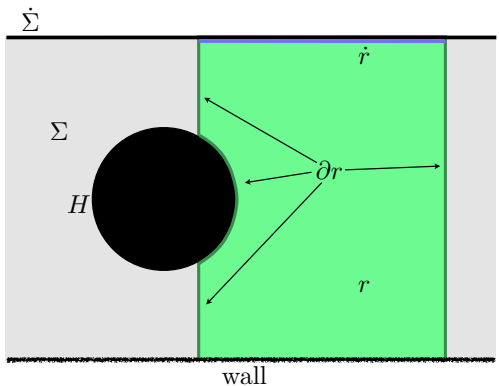}
\caption{\label{fig:defs}
Illustration of the various regions and boundaries defined in section \ref{sec:setup}. $\Sigma$ is a constant-time slice of the bulk spacetime $M$. $\dot\Sigma$ is its conformal boundary, which is a constant-time slice of the boundary spacetime $\dot M$. $H$ is the horizon of black hole in the bulk. The bulk is also bounded at the bottom by a wall of some kind. $r$ (in green) is a region of the bulk. The part of its boundary along $\dot\Sigma$ is denoted $\dot r$. The rest of its boundary, including the part along $H$ but not including the part along the wall, is denoted $\partial r$.
}\end{figure}

We define a \emph{boundary region} to be a codimension-0 subset of $\dot\Sigma$ which is ``nice'', in the sense that neither it nor its complement contains higher-codimension components and its boundary has locally finite area (i.e.\ is not fractal). We denote boundary regions by $A,B,\ldots$; these are assumed to be non-overlapping, but may be adjacent (i.e.\ their boundaries may overlap). We denote $A\cup B$ by $AB$, etc.

We similary define a \emph{bulk region} as a ``nice'' codimension-0 subset of $\Sigma$, and \emph{surface} as a codimension-1 subset. Given a bulk region $r$, we let $\dot r$ denote the part of its boundary that lies along $\dot\Sigma$. We let $\partial r$ denote the part that lies either in the interior of $\Sigma$ or along $H$ (not along $\dot\Sigma$ or a wall). See figure \ref{fig:defs} for an illustration. Applying this rule to all of $\Sigma$, for example, we have $H=\partial\Sigma$. In other words, morally we consider a wall to be ``inside'' $\Sigma$, rather than a boundary of it.

We will be extensively considering unions, intersections, and differences of both boundary and bulk regions in the properties and proofs that follow. In cases of interest, very often two regions share a boundary, so we should be careful how we treat the shared boundary. Also, a little extra notation will go a long way towards simplifying the proofs.

Given two bulk regions $r_1,r_2$ (possibly overlapping), the surface $\partial r_1$ may be divided into four surfaces:
\begin{enumerate}
\item the part that lies inside of $r_2$, whose area we denote $\inte(r_1,r_2)$;
\item the part that lies outside of $r_2$, whose area we denote $\ext(r_1,r_2)$;
\item the part that lies along $\partial r_2$ with $r_1,r_2$ on opposite sides of the shared boundary, whose area we denote $\opp(r_1,r_2)$;
\item the part that lies along $\partial r_2$ with $r_1,r_2$ on the same side, whose area we denote $\sam(r_1,r_2)$.
\end{enumerate}
We thus have
\begin{equation}\label{drdecomp3}
\area(\partial r_1) = \inte(r_1,r_2)+\ext(r_1,r_2)+\opp(r_1,r_2)+\sam(r_1,r_2)\,.
\end{equation}

We define $r_1\cup r_2$ to include any shared boundary where $r_{1,2}$ are on opposite sides (as if $r_1,r_2$ were closed), but $r_1\cap r_2$ to exclude it (as if they were open); similarly, we define $r_1\setminus r_2$ to exclude a shared boundary with $r_{1,2}$ on the same side. We define the union, intersection, and difference of boundary regions the same way. These definitions ensure that the operators preserve the ``niceness'' of the regions.\footnote{These rules correspond to what would happen if the space was latticized, with the regions being sets of lattice points and their boundaries sets of links.} Their surface areas are then given by the following formulas, which will be used repeatedly throughout this paper:
\begin{eqnarray}
\area(\partial(r_1\cup r_2))&=&\ext(r_1,r_2)+\ext(r_2,r_1)+\sam(r_1,r_2) \label{unionlemma} \\
\area(\partial(r_1\cap r_2))&=&\inte(r_1,r_2)+\inte(r_2,r_1)+\sam(r_1,r_2)
\label{intersectlemma} \\
\area(\partial(r_1\setminus r_2))&=&\ext(r_1,r_2)+\inte(r_2,r_1)+\opp(r_1,r_2)\,.\label{differencelemma}
\end{eqnarray}

Finally, we will often use the fact that the dot (conformal boundary) operator commutes with the union, intersection, and difference operators, e.g.\ $\dot{(r_1\cup r_2)}=\dot r_1\cup\dot r_2$.

\subsection{Ryu-Takayanagi formula}\label{RTstatement}

We are now finally ready to state the Ryu-Takayanagi formula \cite{Ryu:2006bv,Ryu:2006ef}. We will do so in a slightly non-standard way, because we wish to emphasize the bulk region that the formula associates to each boundary region. Given a region $A\subseteq \dot\Sigma$, the RT formula gives its entanglement entropy in the state $\rho$ as
\begin{equation}\label{RT}
S(A) = \min_{r\subseteq\Sigma: \dot r=A}\left(\area(\partial r)\right).
\end{equation}
We will denote the minimizer $r(A)$, and define $m(A):=\partial r(A)$, so we have $S(A)=\area(m(A))$, which is the usual statement of the RT formula. In terms of $m(A)$, the condition $\dot r=A$ incorporates both the anchoring condition $\dot m(A)=\partial A$ and the so-called homology condition.

Even in the presence of an ultraviolet cutoff, $S(A)$ can be infinite due to an infrared divergence. We assume that some infrared cutoff has been imposed, so that $S(A)$ is finite for all regions of interest, as is $S_{\rm tot}$.\footnote{There are two issues that make an IR cutoff desirable. First, while one can require $m(A)$ to be locally minimal, it is difficult to define a globally minimal surface when $\area(\partial r)$ is infinite for all suitable $r$. Second, in the presence of translational symmetry of both $M$ and $A$, one may be interested in the ``EE per unit length'', which is most easily defined by first introducing an IR cutoff.} 

Implicit in the definitions of $r(A)$ and $m(A)$ above are the assumptions that the minimizer exists and is unique. The existence of a minimizer is crucial for almost everything we do in this paper. Following common practice, we will more or less it for granted, but let us make a few comments. The existence of a minimal surface with a prescribed boundary (Plateau's problem) has been proven rigorously in various contexts (see for example \cite{MR2455580}), including in hyperbolic spaces with prescribed boundary on the conformal boundary \cite{MR705537}. The main new issue in our case is that $\Sigma$ may be bounded by horizons and walls, and $\partial r$ could ``run off'' to one of these boundaries.  (It won't run off to $\dot\Sigma$, since the metric on $\Sigma$ is large in the vicinity of this boundary.) While this is certainly a possibility, it produces no conflict with the existence of a minimizer, since $r(A)$ can itself be bounded by the wall or horizon (even one that is infinitely far away). It is interesting to note that, even when this occurs, $m(A)$ will still be a stationary point of the area functional. (This is a non-trivial statement because the minimum of a function occurring on the boundary of its domain need not in general be a stationary point.) If $m(A)$ coincides (entirely or in part) with $H$, then this follows from the fact that, being a bifurcate horizon, $H$ is itself an extremal surface. And if $m(A)$ intersects a wall, it will do so perpendicularly, and therefore still be a stationary point of the area. It is important here that the wall is not included as part of $m(A)$.\footnote{A counterexample can be constructed by combining a wall and a horizon, specifically by allowing them to intersect at an obtuse angle. Then a minimal surface that coincides with $H$ will not intersect the wall perpendicularly. We are not aware of an otherwise physically reasonable example where this happens.}

On the other hand, uniqueness of the minimizer is definitely not always the case. For example, it is well known that the globally minimal surface can jump between two locally minimal surfaces as the region $A$ is varied, in analogy to a first-order phase transition, and at the transition point the two minimal surfaces will have equal area (see Property \ref{sec:continuity}). However, as far as we are aware, the minimizer is always \emph{generically} unique: if it is not unique, then after a small change in $A$ it will become unique. For the most part, we will assume uniqueness, because this will substantially simplify both our notation and several of our proofs. However, we will endeavor to point out when this assumption is more than just a convenience, and how it can be relaxed.

Finally, let us make two comments about the proofs that follow: First, there will be little attempt at rigor; rather, our main purpose will be to make explicit the important physical assumptions that stand behind each property. Second, for many of the proofs we will give sketches to illustrate the constructions. These sketches involve one-dimensional boundaries and two-dimensional bulks, with the simplest topologies possible to give the necessary illustration. However, one should keep in mind that, except where otherwise noted, the properties hold irrespective of the dimension and topology of the boundary, bulk, and regions involved.

\section{Continuity of $S$}\label{sec:continuity}

This property states that, if $A_x$ is a continuous one-parameter family of regions, then $S(A_x)$ is a continuous function of $x$.

\emph{Proof:}\footnote{See \cite{Hubeny:2013gta} for an alternative discussion.} Essentially, this property follows from the fact that $S(A)$ is defined by a global minimization, so even if there are competing local minima, the value at the global minimum will be continuous. A more careful argument, which rules out the possibility that a minimum could simply disappear, is the following. We wish to show that $\Delta S:=S(A_{x+\Delta x})-S(A_x)$ goes to 0 as $\Delta x\to0$. We define the region $r'(A_{x+\Delta x})$ such that $\dot r'(A_{x+\Delta x})=A_{x+\Delta x}$, by deforming $r(A_x)$ only in a small neighborhood of $\dot\Sigma$. The difference in area $\Delta S':=\area(\partial r'(A_{x+\Delta x}))-S(A_x)$ goes to 0 as $\Delta x\to0$. Since $S(A_{x+\Delta x})\le\area(\partial r'(A_{x+\Delta x}))$, we have $\Delta S\le\Delta S'$, and so $\lim_{\Delta x\to0}\Delta S\le0$. By deforming $r(A_{x+\Delta x})$, the same argument gives $\lim_{\Delta x\to0}\Delta S\ge0$, so together we have $\lim_{\Delta x\to0}\Delta S=0$. $\Box$ 

Notice that this proof does not imply that $S$ is continuously differentiable. Indeed, the minimizer can switch discontinuously between distinct (typically topologically distinct) local minima, so that $r(A_x)$, $m(A_x)$, and $dS(A_x)/dx$ need not be continuous. For example, for some range of $x$ values we may have two local minima $m_{1,2}(x)$, with areas $S_{1,2}(x)$ respectively, such that $S_1(x)<S_2(x)$ ($S_1(x)>S_2(x)$) for $x<x_c$ ($x>x_c$). Then
\begin{equation}\label{phasetransition}
S(A_x) = \begin{cases}S_1(x)\,,&x<x_c \\ S_2(x)\,,&x>x_c\end{cases}\,.
\end{equation}
Many examples of such ``phase transitions'' are known (see \cite{Hirata:2006jx,Nishioka:2006gr,Klebanov:2007ws,Headrick:2010zt} for early ones).

Is the continuity of $S$ required for consistency, or is this property special to holographic theories? In a general quantum-mechanical system, we cannot usually continuously vary the subsystem $A$, so in asking this question we will restrict ourselves to the context of quantum field theories. It is generally believed that, in a field theory with a finite number of fields, as long as $A$ is bounded, there should be no phase transitions in $S(A)$. Phase transitions are certainly possible either in the thermodynamic limit or in infinite volume, but we are not aware either of an example where $S(A)$ jumps as a function of $A$, or of an argument that it cannot do so.

Therefore, it is interesting to ask what we can learn about the structure of the reduced density matrix $\rho_A$ in holographic theories from the fact that $S(A)$ is defined by a global minimization, and hence is continuous across phase transitions. In the example above, it seems reasonable to infer that the reduced density matrix $\rho_{A_x}$ includes sub-ensembles $\rho_{1,2}(x)$, with entropies $S_{1,2}(x)$ respectively, both of which are present over the whole range of $x$ but exchange dominance at $x=x_c$. In other words we have, at least roughly,
\begin{equation}\label{roughly}
\rho_{A}=p_1\rho_1\oplus p_2\rho_2
\end{equation}
(with $p_i>0$, $p_1+p_2=1$; the $p_i$ also depend on $x$, but for clarity we've dropped all the $x$-dependences). \eqref{roughly} implies
\begin{equation}\label{roughent}
S(A) = p_1S_1+p_2S_2-p_1\ln p_1-p_2\ln p_2\,.
\end{equation}
A similar situation occurs in the microcanonical ensemble, where there may be a competition between different macrostates with the same energy. In the simple case where there are just two macrostates, \eqref{roughly} and \eqref{roughent} apply; since all microstates are weighted equally, $p_i=e^{S_i}/(e^{S_1}+e^{S_2})$, and, in the thermodynamic limit, $S=\max(S_1,S_2)$ (where all quantities are functions of the energy). The entropy is continuous but has a discontinuous second derivative as a function of the energy, just as for the EE as a function of $x$. Unlike for the EE, however, it is the macrostate with the \emph{largest} entropy that dominates. From this point of view, the behavior of the holographic EE seems strange. In order for the sub-ensemble with the \emph{smallest} entropy to win, it must be that each state in sub-ensemble $\rho_i$ is weighted inversely with $e^{S_i}$. For example, a simple possibility that yields \eqref{phasetransition} in the thermodynamic limit is
\begin{equation}\label{probabilities}
p_i = \frac{e^{-S_i}}{e^{-S_1}+e^{-S_2}}\,.
\end{equation}

\section{Properties of the map $r$}\label{sec:mapr}

As mentioned in the Introduction, one interesting but perhaps underappreciated feature of the Ryu-Takayanagi formula is that it associates, in a canonical and geometrically natural way, a bulk region $A$ to each boundary region $r(A)$. In this section we will give four properties that this map obeys. While some of the these properties have been mentioned in passing in previous work, we are not aware of a systematic treatment. In subsection \ref{interpretation}, we will then give a possible physical interpretation of this map, in view of these properties.

\subsection{$r(\emptyset)=\emptyset$}

This immediately implies $m(\emptyset)=\emptyset$, $S(\emptyset)=0$. The latter fact is required for consistency.

\emph{Proof:} Since $r=\emptyset$ is certainly allowed, since it obeys $\dot\emptyset=\emptyset$. Furthermore, it is the minimizer: Since $\Sigma$ is connected and equipped with a positive-definite metric, the only regions such that $\area(\partial r)=0$ are $r=\emptyset$ and possibly $r=\Sigma$; but the latter does not obey $\dot r=\emptyset$. $\Box$

\subsection{$r(\dot \Sigma)=\Sigma$}\label{r(dotSigma)}

This immediately implies $m(\dot \Sigma)=H$, $S(\dot \Sigma)=\area(H)=S_{\rm tot}$. The latter fact is clearly required for consistency, given that, by the Bekenstein-Hawking formula, $S_{\rm tot}$ is the entropy of $\rho$.

\begin{figure}[tbp]
\centering
\includegraphics[width=0.9\textwidth]{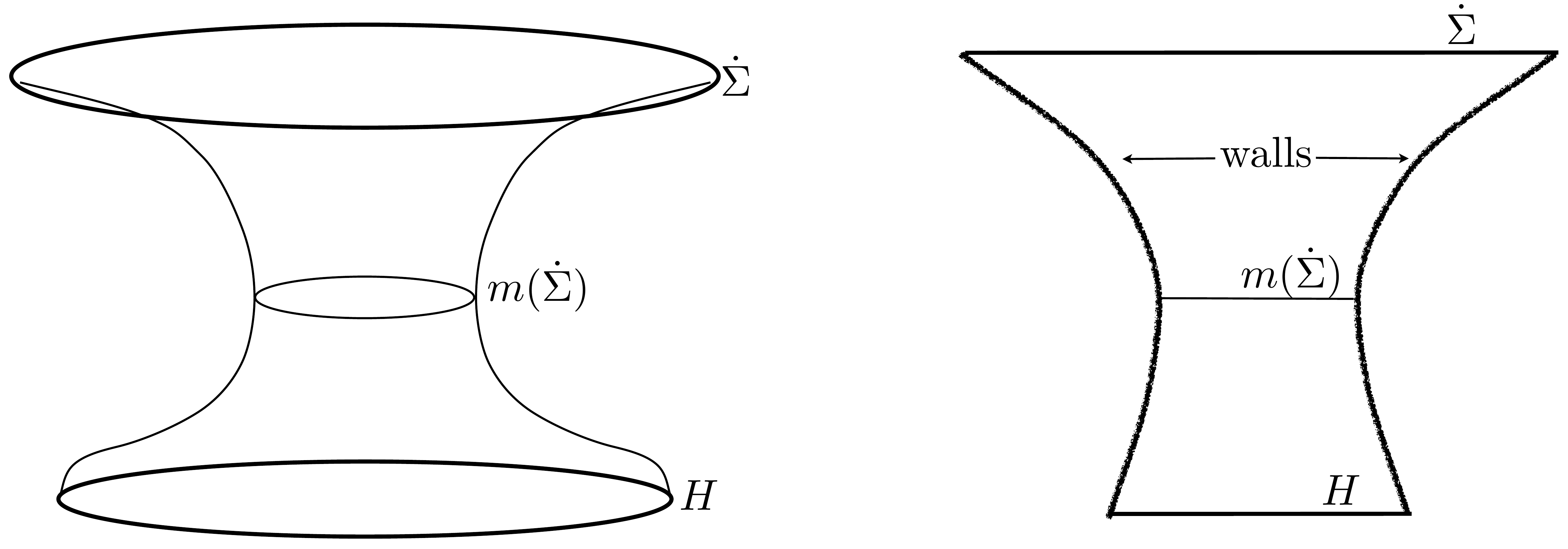}
\caption{\label{fig:rdotSigma}
Left: Situation excluded by Property \ref{r(dotSigma)}, in which there exists a minimal surface homologous to $\dot \Sigma$ with smaller area than $H$, that would therefore be $m(\dot \Sigma)$. (The proof actually excludes the existence of even a local minimal surface homologous to $\dot \Sigma$ other than $H$.) Right: Counterexample to Property \ref{r(dotSigma)}: The walls on the left and right bend in to create a minimal surface homologous to $\dot \Sigma$, with smaller area than $H$. Presumably such behavior for walls is unphysical.}\end{figure}

We will not give a complete proof of this property. Essentially, what we wish to exclude is the existence of a minimal surface other than $H$ that could serve as $\partial r$ and possibly have smaller area than $H$ (see figure \ref{fig:rdotSigma}). (Such a surface would in essence be a traversable wormhole, so the statement is akin to the topological censorship theorem \cite{Friedman:1993ty}.) To do so, we will assume that there are no walls and that the metric on $\Sigma$ is smooth. We will also appeal to the Einstein equation and the null energy condition. With arbitrary placement of walls, or a singular metric on $\Sigma$, it is easy to construct a counterexample (see for example the right side of figure \ref{fig:rdotSigma}). However, we are not aware of one that is otherwise physically reasonable. It seems likely that, with appropriate physical conditions on walls and singularities, one could prove the property even in their presence.

\emph{Proof:} We will first assume that $H$ does not intersect $\dot \Sigma$; in the next paragraph we will relax this assumption. For any region $r$ such that $\dot r=\dot \Sigma$ and $\area(m)$ is finite, where $m:=\partial r$, $m$ is necessarily closed. We will show that, if $m$ is also minimal, then $m=H$. Send out a congruence of future-directed null geodesics orthogonally from $m$, in the direction of $r$. Since $m$ is minimal, this congruence starts out with zero expansion; by a standard application of the Einstein and Raychaudhuri equations and null energy condition, the expansion cannot become positive. Consider the intersection of the congruence with a constant-time slice a short time later than $\Sigma$; by transporting this surface along the Killing vector back to $\Sigma$, we obtain a surface $m'$. Since the expansion is not positive, $\area(m')\le\area(m)$. Now, for points on $H$, the null congruence simply follows the horizon. (Recall that $H$ is the bifurcation surface of the horizon.) On the other hand, since the Killing vector is null only on the horizon, and timelike elsewhere, if $m\neq H$ then $m'\neq m$. (Note that proper subsets of $H$ are not homologous to $\dot \Sigma$, so if $m\neq H$ then some points of $m$ are not in $H$.) Since $m$ is minimal, any small variation increases its area, so $\area(m')>\area(m)$, and we have arrived at a contradiction.

Now suppose $H$ does intersect $\dot \Sigma$; call the intersection $\dot H$. Necessarily $\dot m=\dot H$. Since $m$ is minimal, its area increases under any small variation that fixes $\dot m$, where it is anchored. So, to run the argument from the previous paragraph, we only need to show that $\dot m'=\dot m$. If $m$ is minimal then it hits $\dot \Sigma$ perpendicularly (at least in the limit that the UV cutoff is removed), so along the boundary spacetime, the null congruence coincides with the horizon, hence $\dot m'=\dot m$. $\Box$

Being the bifurcation surface of a Killing horizon, $H$ is necessarily extremal. However, a bifurcation surface can be minimal (as for a black-hole horizon), maximal (as for a cosmological horizon), or neither (as for a Rindler horizon). An interesting corollary of this property is that $H$ must be minimal (presuming, as we do throughout this paper, that a minimal surface exists).

\subsection{$r(A)\subseteq r(AB)$}\label{monotonicity}

If we do not assume uniqueness of the minimizers, then the precise statement is that $r(A)$, $r(AB)$ can be chosen so that $r(A)\subseteq r(AB)$.

If $B=\emptyset$ then the property is trivial, so we will henceforth assume $B\neq\emptyset$. Necessarily $r(A)\neq r(AB)$, so the property states that $r(A)\subset r(AB)$, in other words that the map $r$ is strictly monotonic.

\begin{figure}[tbp]
\centering
\includegraphics[width=0.8\textwidth]{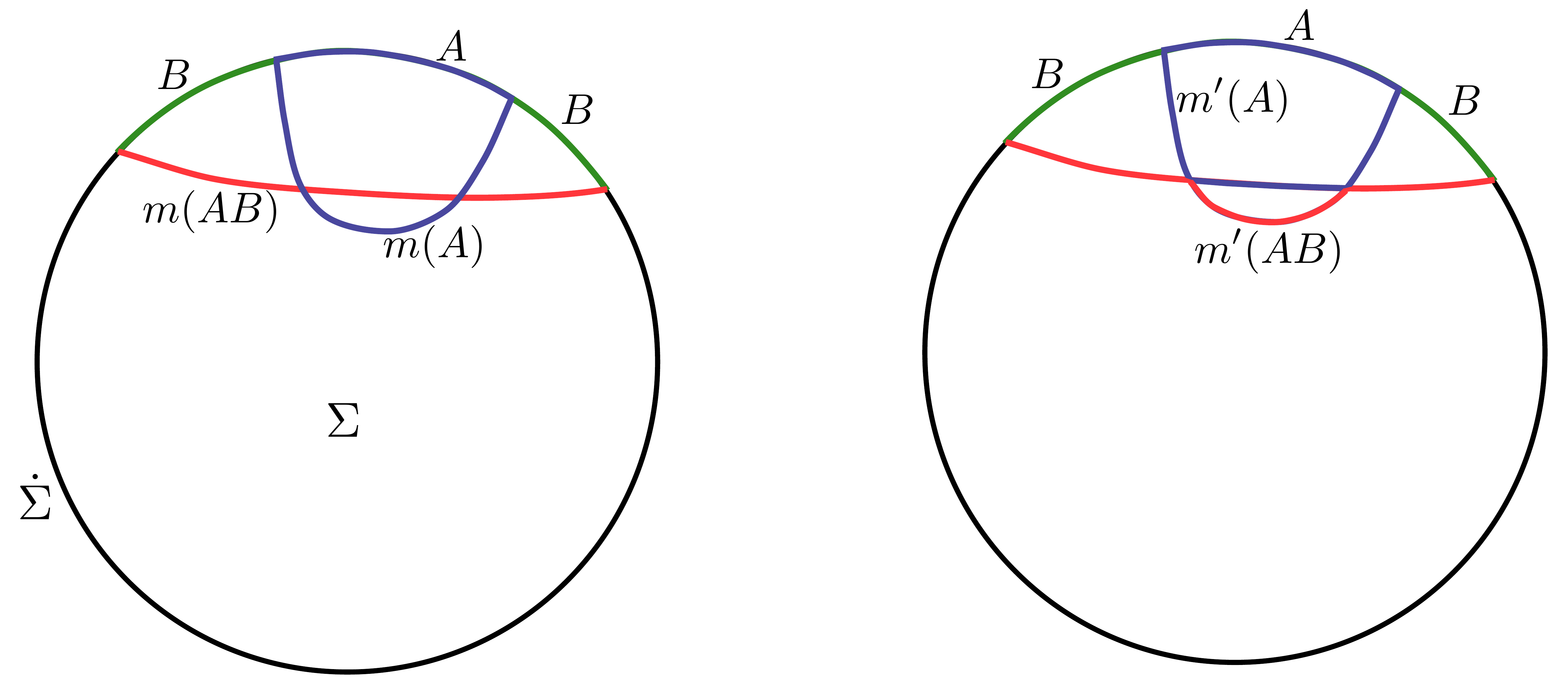}
\caption{\label{fig:monotonic}
Left: Illustration of the situation excluded by Property \ref{monotonicity}. $A$ and $m(A)$ are in blue, $B$  is in green, and $m(AB)$ is in red. Right: Surfaces $m'(A):=\partial r'(A)$ (blue) and $m'(AB):=\partial r'(AB)$ (red) used in the proof of Property \ref{monotonicity}, for the surfaces shown on the left.}
\end{figure}

\emph{Proof:} We proceed by contradiction. Define two new regions
\begin{equation}
r'(A):=r(A)\cap r(AB)\,,\qquad r'(AB):=r(A)\cup r(AB)\,,
\end{equation}
and their corresponding boundary areas, $S'(A):=\partial r'(A)$, $S'(AB)=\partial r'(AB)$. (See figure \ref{fig:monotonic}.) The regions satisfy $\dot r'(A)=A$, $\dot r'(AB)=AB$, but are distinct from $r(A),r(AB)$.\footnote{Since $r(A)\neq r(AB)$, if $r(A)\not\subset r(AB)$, some part of $r(A)$ lies outside of $r(AB)$. Hence $r'(A)\neq r(A)$, $r'(AB)\neq r(AB)$.} So if we assume uniqueness of the minimizer then we must have
\begin{equation}\label{fromuniqueness}
S'(A)>S(A)\,,\qquad S'(AB)>S(AB)\,.
\end{equation}
On the other hand, using \eqref{drdecomp3}, \eqref{unionlemma}, \eqref{intersectlemma}, we have
\begin{equation}
S'(A)-S(A)+S'(AB)-S(AB)=-2\opp(r(A),r(AB))\le0\,,
\end{equation}
which is a contradiction. If we don't assume uniqueness, then the inequalities in \eqref{fromuniqueness} become non-strict, so there is the possibility that
\begin{equation}\label{equality}
S'(A)=S(A)\,,\qquad S'(AB)=S(AB)\,,\qquad \opp(r(A),r(AB))=0\,.
\end{equation}
But in that case, since their areas equal those of minimizers, $r'(A),r'(AB)$ must themselves also be minimizers, and they certainly obey $r'(A)\subset r'(AB)$. $\Box$

This property is required for consistency in the case when $M$ is a subset of a larger static spacetime $M'$ (for example, the Poincar\'e patch or a Rindler wedge inside AdS) and $\Sigma$ is a subset of a constant-time slice $\Sigma'$ of $M'$. If it true that $M$ represents the state of the field theory on $\dot M$, then the EE of any region $A\subseteq\dot\Sigma$ should be computable from $\Sigma$, without knowing the larger space $\Sigma'$, in other words we need $r(A)\subseteq\Sigma$.

\subsection{$r(A)\cap r(B)=\emptyset$}\label{nonoverlapping}

\begin{figure}[tbp]
\centering
\includegraphics[width=0.8\textwidth]{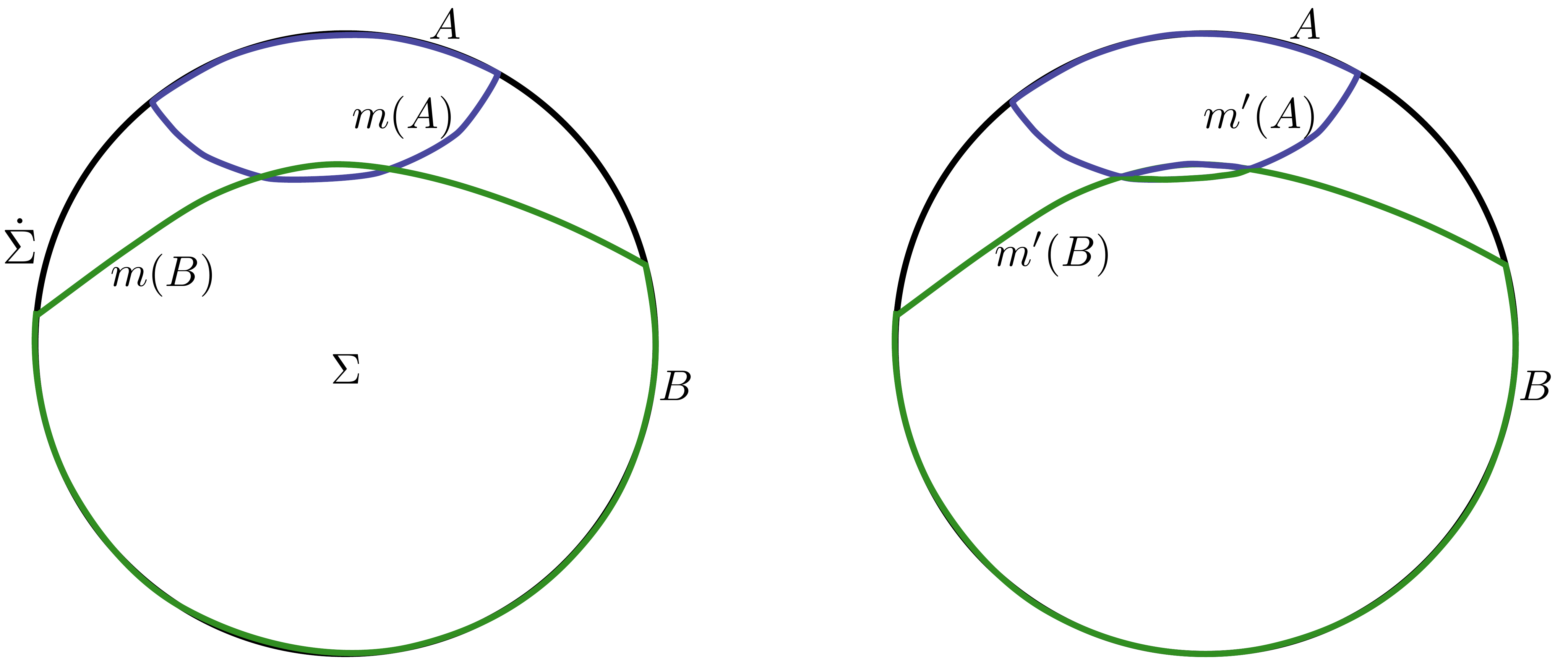}
\caption{\label{fig:nonoverlapping}
Left: Illustration of the situation excluded by Property \ref{nonoverlapping}. $A$ and $m(A)$ are in blue, while $B$ and $m(B)$ are in green. Right: Surfaces $m'(A):=\partial r'(A)$ (blue) and $m'(B):=\partial r'(B)$ (green) used in the proof of Property \ref{nonoverlapping}, for the surfaces shown on the left.}
\end{figure}

\emph{Proof:} Same argument as for Property \ref{monotonicity}, but with $r'(A):=r(A)\setminus r(B)$, $r'(B):=r(B)\setminus r(A)$ (see figure \ref{fig:nonoverlapping}). $\Box$

\subsection{Interpretation}\label{interpretation}

The four properties of the map $r$ listed above, together with the fact that, by definition, $\dot r(A)=A$, strongly suggest that the bulk region $r(A)$ should represent some natural object in the field theory associated to the region $A$. The additional fact that $S(A)=\area(\partial r(A))$ suggests further that this object is the reduced density matrix $\rho_A$, whose von Neumann entropy is $S(A)$. In other words, if you know $r(A)$---including its topology and the full classical field configuration on it---then you know $\rho_A$, even if you don't know anything about the rest of the bulk $\Sigma$. (This statement is similar to one of the proposals by Czech et al. in \cite{Czech:2012bh}.)

Of course, such a statement must be understood in a large-$N$ sense. (In this paragraph, for concreteness, we adopt the language of a gauge/string duality.) Indeed, even the usual statement that the classical field configuration on $\Sigma$ represents the state $\rho$ on $\dot\Sigma$ is only true at leading order in $1/N$. For example, if $\dot\Sigma$ is a sphere and the field theory is a CFT, then the vacuum and a thermal state below the Hawking-Page temperature are represented by the same classical spacetime, namely global AdS. Knowing $\Sigma$, one can directly read off the leading terms in the one-point functions of local single-trace operators from the asymptotic behavior of the fields near $\dot\Sigma$. However, higher-order terms in one-point functions and connected higher-point functions require knowledge of the quantum state of the fields on $\Sigma$ (to say nothing of corrections that are non-perturbative in $1/N$). The same is true for one-point functions of local single-trace operators on $A$, given knowledge of $r(A)$. (It also holds, albeit in a trivial way, for non-local single-trace operators such as connected spacelike Wilson loops. Their one-point functions are of the form $e^{-a_{\rm ws}/\alpha'}$, where $a_{\rm ws}$ is the string-frame area of the minimal worldsheet ending on the boundary, which vanishes in the limit of Einstein gravity, $\alpha'\to0$.\footnote{Note that, even if the Wilson loop lies entirely within $A$, the minimal worldsheet will not in general lie within $r(A)$. To construct a counterexample, fix a Wilson loop and let $A$ be a tubular neighborhood of it that is much smaller in radius than the size of the Wilson loop; $r(A)$ will be a small half-tube very close to $\dot M$, and will certainly not contain the minimal worldsheet ending on the Wilson loop.})

Note that it is \emph{not} in general true that $r(AB)=r(A)\cup r(B)$ (nor the weaker statement $r(A^c)=r(A)^c$). (In section \ref{sec:inequalities}, we will give sufficient conditions for this equality to hold.) What, then, does the remainder $r(AB)\setminus(r(A)\cup r(B))$ represent? In general, knowledge of $\rho_A$ and $\rho_B$ does not fix $\rho_{AB}$. So presumably the rest of $r(AB)$ encodes how $\rho_A$ and $\rho_B$ fit inside $\rho_{AB}$.\footnote{Similar speculations appeared previously in \cite{Czech:2012bh}.}

Of course it would be useful to make the above speculations more precise. The map $r$ will play an essential role in the properties and examples we will study in the next section, and in turn they will help us build intuition about its physical meaning.

\section{Inequalities and their saturation}\label{sec:inequalities}

In this section we discuss inequalities obeyed by the entanglement entropy, as calculated by the Ryu-Takayanagi formula. The proofs of the inequalities were previously published, with two exceptions. First, we fill a small gap in the published proofs of strong subadditivity and monogamy of mutual information. Second, the last inequality, Property \ref{reflection}, is new.

For several of the inequalities, we will also give necessary and sufficient conditions for their saturation, in terms of the relevant bulk regions and surfaces. The sufficiency of the conditions is obvious, so we will only give proofs of their necessity. It will be convenient to assume uniqueness of the minimizer in these proofs. If one doesn't assume uniqueness, then the correct statement is that the relevant bulk regions $r(A)$ etc.\ can be chosen such that the given condition holds. In each case, we will also discuss the interpretation of the saturation from the field-theory viewpoint, which turns out to reveal quite a bit about the structure of reduced density matrices in holographic theories. The following point will play a crucial role: Since the RT formula only gives the order $1/G_{\rm N}$ part of the entanglement entropy, by ``saturation'' in this context we actually mean ``saturation at order $1/G_{\rm N}$''. In fact, as we will discuss, we do not expect \emph{any} of these inequalities to be exactly saturated, except in trivial cases. For definiteness, it will be convenient to adopt the language of large-$N$ gauge theories (for example, ``gluon'' and ``glueball'' degrees of freedom). However, very similar statements can be made about other holographic theories, such as two-dimensional CFTs where $1/G_{\rm N}\sim c$.

The inequalities in this section are logically related to each other in various ways. For example, \ref{SSA1} implies \ref{subadditivity}, and \ref{SSA2} implies \ref{ArakiLieb}. However, it is useful to list them separately in order to clarify the exposition, especially as regards the conditions for saturation.

\subsection{$S(A)\ge0$}

\emph{Proof:} Obvious from the definition. $\Box$

This is a general property of entropy in any quantum system, and is therefore required for consistency.

\subsubsection{Saturation}\label{positivitysaturation}

\emph{Condition:} $S(A)=0$ if and only if either $A=\emptyset$ or $A=\dot\Sigma$ and $S_{\rm tot}=0$.

\emph{Proof:} Given that $\Sigma$ is assumed to be connected and carries a positive-definite metric, any non-empty proper subset $r\subset\Sigma$ has $\partial r>0$. So if $S(A)=0$ then either $r(A)=\emptyset$ or $r(A)=\Sigma$. The former implies $A=\emptyset$, while the latter implies $A=\dot\Sigma$. $\Box$

At first glance this statement seems to merely say that non-trivial regions are always entangled, which we would expect in any field theory. However, it actually says that the EE is always of order $1/G_{\rm N}\sim N^2$. The physical interpretation is that, in any state that can be described holographically, the entanglement across any entangling surface involves the gluonic (i.e.\ non-gauge-invariant) degrees of freedom. This is a statement about short-distance correlation; in the next subsection, we will see that it is not necessarily the case for measures of long-distance correlation.

\subsection{$S(AB)\le S(A)+S(B)$}\label{subadditivity}

\emph{Proof:} Define $r'(AB):=r(A)\cup r(B)$, $S'(AB):=\area(\partial r'(AB))$. This satisfies $\dot r'(AB)=AB$, so $S(AB)\le S'(AB)$. From \eqref{unionlemma} we have
\begin{equation}\label{SAproof1}
S'(AB) = \ext(r(A),r(B))+\ext(r(B),r(A))\,,
\end{equation}
while from \eqref{drdecomp3} we have
\begin{equation}\label{SAproof2}
S(A)+S(B) = \ext(r(A),r(B))+\ext(r(B),r(A)) + 2\opp(r(A),r(B))\,,
\end{equation}
where we have used the fact that, by Property \ref{nonoverlapping}, $\sam(r(A),r(B))=\inte(r(A),r(B))=\inte(r(B),r(A))=0$. So $S'(AB)\le S(A)+S(B)$. $\Box$

This property is called \emph{subadditivity}. It is a general property of entropy in any quantum system, and is therefore required for consistency. The difference between the two sides defines the \emph{mutual information},
\begin{equation}
I(A:B):=S(A)+S(B)-S(AB)\,,
\end{equation}
which quantifies the total amount of correlation between $A$ and $B$, including both classical correlation and entanglement. For example, a pair of bits in $A,B$ in the maximally entangled state $\frac12(\ket{00}+\ket{11})(\bra{00}+\bra{11})$ contributes $2\ln2$ to $I(A:B)$, while a pair in the maximally classically correlated state $\frac12(\ket{00}\bra{00}+\ket{11}\bra{11})$ contributes $\ln2$.

\subsubsection{Saturation}\label{subadditivitysaturation}

\begin{figure}[tbp]
\centering
\includegraphics[width=0.75\textwidth]{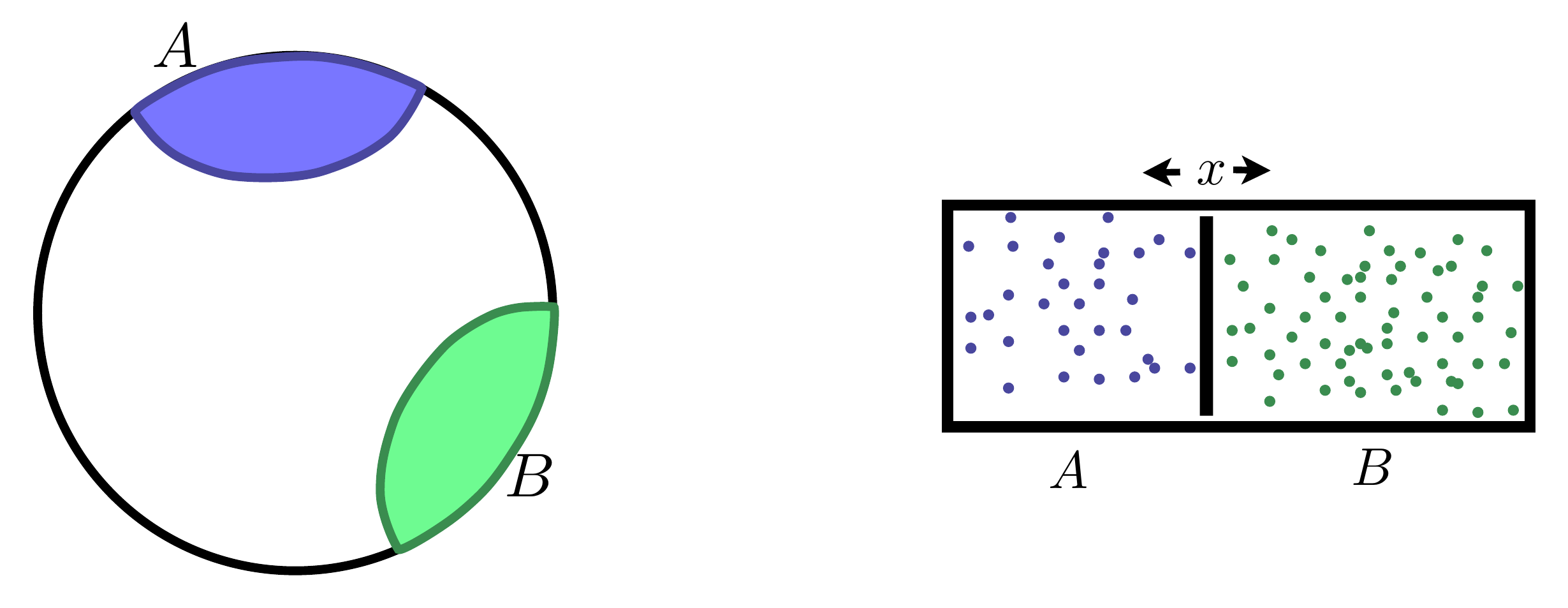}
\caption{\label{fig:subsaturation}
Left: Illustration of a case where the subadditivity inequality is saturated. Right: Example of two thermodynamic systems coupled via a macroscopic variable: Two species of gas in a box separated by a movable piston. $A,B$ represent the states of the two gases respectively. As discussed in the main text, such a system is closely analogous to the state of regions in a holographic field theory such that subadditivity is saturated, as on the left side.}
\end{figure}

\emph{Condition:} $S(AB)=S(A)+S(B)$ if and only if $r(AB)=r(A)\cup r(B)$ and $m(AB)=m(A)\cup m(B)$ (see figure \ref{fig:subsaturation}, left side).

\emph{Proof:} Looking at the proof of subadditivity, we see that its saturation implies
\begin{equation}
S(AB)=S'(AB)=S(A)+S(B)\,.
\end{equation}
The first equality, along with uniqueness of the minimizer, implies $r(AB)=r'(AB)$. It then follows from the second that $\opp(r(A),r(B))=0$, hence $m(AB)=m(A)\cup m(B)$. $\Box$

Subadditivity is saturated whenever the regions $A,B$ are far enough apart, relative to their sizes and any scales in the background, for the preferred minimal surface not to join them.\footnote{The discussion in this paragraph and the next one follows \cite{Headrick:2010zt}.} If we fix their sizes and shapes and vary their separation continuously, there is usually a first-order phase transition at some separation, in which $I(A:B)$ goes from being zero to non-zero, continuously but with a discontinuous first derivative. The simplest example is the vacuum of a CFT on the line in $1+1$ dimensions, with $A,B$ each a single interval, where the phase transition occurs when the cross-ratio of their four endpoints is 1/2.

It is interesting to ask what the field-theory interpretation of this situation is. In a general quantum system, $I(A:B)$ vanishes exactly if and only if $\rho_{AB}=\rho_A\otimes\rho_B$. This implies that the connected two-point functions between any operator in $A$ and any operator in $B$ vanish, which would be rather surprising in a field theory. However, in the holographic case, $I(A:B)$ does not strictly vanish, but rather is of order $G_{\rm N}^0\sim N^0$. (There must also exist non-perturbative corrections to the EE, which smooth out the phase transition at finite $N$.) This indicates that the correlations are being carried only by gauge-invariant degrees of freedom (glueballs, etc.); on the other hand, when $I(A:B)$ is of order $1/G_{\rm N}\sim N^2$, the amount of correlation is so large that it must be being carried by the colored degrees of freedom (gluons, etc.).

The same phenomenon occurs for any two thermodynamic systems that are coupled to each other only via macroscopic observables. For example, let $A,B$ be two species of gas in a box separated by a movable piston (see figure \ref{fig:subsaturation}, right side). The states of $A,B$ will be correlated due to fluctuations in the position of the piston. However, in the thermodynamic limit these fluctuations are small, and therefore $I(A:B)$ is small. More precisely, if there are of order $M$ molecules of each gas, then the entropies $S(A)$, $S(B)$, $S(AB)$ are of order $M$, but the fluctuations in the position of the piston are of order $1/\sqrt{M}$ and the mutual information is of order 1. To see this, write the state $\rho_{AB}$ as a direct sum of states with definite values of the macroscopic variable $x$; since $A,B$ are coupled only via $x$, each state in the direct sum is a tensor product:
\begin{equation}\label{macrocoupled}
\rho_{AB}=\frac1Z\bigoplus_xe^{-F(x)}\rho_A(x)\otimes\rho_B(x)\,,
\end{equation}
where $F(x)$ is the free energy for fixed $x$. Since the systems are macroscopic, $F(x)$ is of order $M$. A short calculation then shows
\begin{equation}
I(A:B)=\frac1Z\int dx\left(F(x)-\ln Z\right)e^{-F(x)}\,;
\end{equation}
this is of order 1, since in the leading saddle-point approximation $\ln Z\approx F(x_1)$, where $x_1$ is the equilibrium value of $x$ (the minimum of $F(x)$). Similarly, $\ev{(x-x_1)^2}$ is of order $1/M$. (On the other hand, if the two gases are allowed to mingle, then the mutual information will be extensive in $M$.)

The large-$N$ limit of a gauge theory is a thermodynamic limit, where the ``macroscopic observables'' are the gauge-invariant operators. In a holographic theory, the corresponding degrees of freedom are the bulk fields. So, by analogy to the box of gas, we conclude that when the regions $A,B$ are sufficiently far separated, their mutual information can be understood in terms of fluctuations of the bulk fields. While from the field-theory point of view these are statistical fluctuations, from the bulk point of view they are quantum fluctuations, and indeed are of order $\sqrt{G_{\rm N}}\sim 1/N$, as expected from the above reasoning, with $M\sim N^2$. (See also \cite{Faulkner:2013ana,Barrella:2013wja}.)

The view of far-separated regions as thermodynamic systems coupled via macroscopic observables also demystifies a puzzling feature of their R\'enyi entropies. Replica-trick calculations of R\'enyi entropies for the example mentioned above (two disjoint intervals in a two-dimensional CFT on the line in the vacuum) revealed that, even when $I(A:B)$ is of order 1, the mutual R\'enyi information (MRI) $I_\alpha(A:B):=S_\alpha(A)+S_\alpha(B)-S_\alpha(AB)$ ($\alpha\neq1$) is of order $1/G_{\rm N}\sim c$ \cite{Headrick:2010zt}. At first sight, this large MRI is rather surprising: given that a strictly vanishing mutual information implies a vanishing MRI, one might have expected that when the former is small the latter is also small. However, since the MRI is known not to be a good measure of correlation---for example, it is not positive or monotonic under inclusion---the physical significance of this large value was not clear.

In fact, this behavior is not at all surprising when we view $A,B$ as macroscopically coupled systems. Given the ensemble \eqref{macrocoupled}, a short calculation shows that $I_\alpha(A:B)$ is of order $M$ for $\alpha\neq1$. The reason is that changing $\alpha$ away from 1 changes the saddle-point value of $x$ by an amount of order 1, and so effectively changes the macroscopic state (unlike the small fluctuations in $x$ that lead to the order-1 value of $I(A:B)$). For example, for the box of gas, changing $\alpha$ effectively changes the temperature; if the two gases have different equations of state, say, then the equilibrium position of the piston will shift as a result.

Thus the state \eqref{macrocoupled}, representing thermodyanic systems coupled via macroscopic variables, reproduces several qualitative features of the reduced density matrix for far-separated regions in a holographic theory. However, we should also note an important difference: The state \eqref{macrocoupled} is separable, meaning that there are only classical correlations but no entanglement between $A$ and $B$, while it is expected that regions in quantum field theories always have some entanglement between them \cite{SummersWerner87}. Thus the state $\rho_{AB}$ in the holographic case is likely more complicated than \eqref{macrocoupled}.

\subsection{$S(A)\le S(AB)+S(B)$}\label{ArakiLieb}

\emph{Proof:} Same strategy as for proof of subadditivity, with $r'(A):=r(AB)\setminus r(B)$. $\Box$

This is called the \emph{Araki-Lieb} (AL) or \emph{triangle inequality}. It is often written $|S(A)-S(B)|\le S(AB)$. It generalizes the statement that, if $\rho_{AB}$ is pure, then $S(A)=S(B)$. It is a general property of entropy in any quantum system, and is therefore required for consistency.

In analogy to the mutual information, we can define the difference between the two sides as the \emph{intrinsic entropy}:
\begin{equation}
J(B,A):=S(AB)+S(B)-S(A)\,,
\end{equation}
which quantifies how much of the entropy in $B$ is \emph{not} due to entanglement with $A$.\footnote{If the full system is in a pure state, then $J(B,A)=I(B:(AB)^c)$.} For example, consider a bit of $B$ that, after tracing over $A$, is in the maximally mixed state $\frac12(\ket{0}\bra{0}+\ket{1}\bra{1})$, and therefore contributes $\ln 2$ to $S(B)$. If, before tracing over $A$, the bit is uncorrelated with $A$, then it contributes $2\ln2$ to $J(B,A)$; on the other hand, if it is maximally classically correlated with some bit in $A$, then it contributes only $\ln2$, while if it is maximally entangled then it does not contribute at all.

\subsubsection{Saturation}\label{ArakiLiebsaturation}

\begin{figure}[tbp]
\centering
\includegraphics[width=\textwidth]{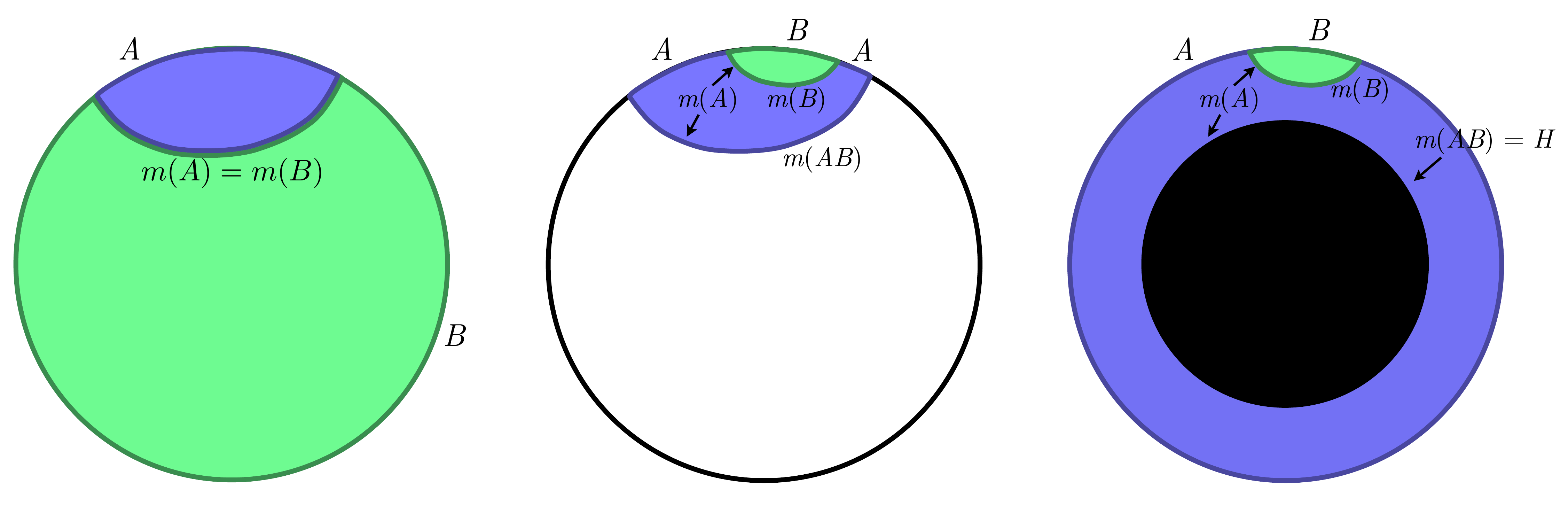}
\caption{\label{fig:ALsaturation}
Illustration of cases where the AL inequality is saturated. In the left case, $B=A^c$ and $S(AB)=S_{\rm tot}=0$; therefore, for any choice of $A$, AL must be saturated and $r(B)=r(A)^c$. In the center and right cases, $S(AB)\neq0$ (in the center, $B\neq A^c$; on the right, $B=A^c$ but $S_{\rm tot}\neq0$); therefore, the fact that AL is saturated depends on the particular arrangement of $A,B$.}
\end{figure}

\emph{Condition:} $S(A)=S(AB)+S(B)$ if and only if $r(A)=r(AB)\setminus r(B)$ and $m(A)=m(AB)\cup m(B)$. (See figure \ref{fig:ALsaturation}.)

\emph{Proof:} Similar to saturation of subadditivity (except the condition $\opp(r(AB),r(B))=0$ is replaced by the condition $\sam(r(AB),r(B))=0$). $\Box$

An important special case is when $S(AB)=0$, in which case AL is necessarily saturated. By property \ref{positivitysaturation}, this can only happen when $B=A^c$ and $S_{\rm tot}=0$. In this case, $r(B)=r(A)^c$; see figure \ref{fig:ALsaturation}, left side. In particular, if the full system is in a strictly pure state, such as the vacuum, and $B=A^c$, then AL is necessarily exactly saturated. On the other hand, if the entropy of the full system is of order 1 (for example, in a thermal state below the Hawking-Page transition), then we would expect AL to only be saturated at order $1/G_{\rm N}$.

AL can still be saturated even when $AB$ is not pure, either because $B\neq A^c$ (figure \ref{fig:ALsaturation}, middle) or because $S_{\rm tot}\neq0$ (i.e.\ the bulk contains a horizon; figure \ref{fig:ALsaturation}, right side), or both. Saturation requires $B$ to be surrounded by $A$, i.e.\ $\partial B\subseteq\partial A$; otherwise we cannot have $r(A)=r(AB)\setminus r(B)$. Roughly speaking, $J(B,A)$ vanishes when $B$ is sufficiently small compared to $A$ and other relevant scales in the background. If we tune the relative sizes of the two regions, a first-order phase transition can occur in which $J(B,A)$ goes from being zero to non-zero, continuously but with a discontinuous first derivative.\footnote{Examples of saturation of AL and the accompanying phase transition, in cases where $B=A^c$ and the bulk contains a horizon, were studied in \cite{Hubeny:2013gta}.} In these cases, we would expect the saturation to occur only at order $1/G_{\rm N}$. There must also be non-perturbative corrections that smooth out the phase transition at finite $G_{\rm N}$, i.e.\ finite $N$.

The saturation of AL has intriguing implications from the field-theory viewpoint. We should first ask what its saturation implies for a general quantum system. Since the intrinsic entropy quantifies how much of the entropy of $B$ is \emph{not} due to entanglement with $A$, we would expect that, when it vanishes, all of $B$ is maximally entangled with all or part of $A$; by the monogamy of entanglement, the rest of $A$ must then be uncorrelated with both $B$ and the first part of $A$. This intuition is confirmed by the following theorem \cite{ZhangWu}: $J(B,A)$ vanishes exactly if and only if the $A$ Hilbert space can be decomposed into two factors, $\mathcal{H}_A=\mathcal{H}_{A_1}\otimes\mathcal{H}_{A_2}$, such that
\begin{equation}
\rho_{AB}=\rho_{A_1}\otimes\rho_{A_2B}\,,
\end{equation}
where $\rho_{A_2B}$ is pure. In other words, the degrees of freedom of $A$ can be divided into two uncorrelated sets, those that carry all of the entanglement with $B$ and those that carry all of the full system's entropy.

Returning to the saturation of AL in the holographic context, we might be tempted to conclude from the above theorem that, again, the $A$ Hilbert space can be decomposed such that $\rho_{AB}=\rho_{A_1}\otimes\rho_{A_2B}$, with $\rho_{A_2B}$ pure. In other words, the degrees of freedom of $A$ can be divided into two uncorrelated sets, one of which is maximally entangled with $B$ and the other of which carries all of the entropy of $AB$. It is clear that such a division \emph{cannot} be only according to geometric regions, since such regions would never be uncorrelated; therefore it must somehow be among the gluonic degrees of freedom. The idea that the gluonic degrees of freedom living at a given point in space can be divided into two uncorrelated sets is quite surprising.

However, the situation is actually more complicated, since in a holographic system we don't expect $J(B,A)$ to vanish exactly, but rather to be of order 1 (except when $B=A^c$ and the full system is in an exactly pure state, in which case the above decomposition is trivial). One simple kind of state consistent with $S(A),S(B),S(AB)=O(1/G_{\rm N})$ and $J(B,A)=O(1)$ is, again, $\rho_{AB}=\rho_{A_1}\otimes\rho_{A_2B}$, with $S(A_1)=O(1/G_{\rm N})$ and $J(B,A_2)=O(1)$ (again, $A_{1,2}$ are not regions); in other words, $B$ is almost but not entirely entangled with part of $A$. However, it seems unlikely that the degrees of freedom in an interacting field theory could ever admit a decomposition into exactly uncorrelated subsets. A more realistic model would include mixtures of such states, analogous to \eqref{macrocoupled}. Thus, we decompose $\mathcal{H}_A$ into a direct sum of products, $\mathcal{H}_A=\bigoplus_i\mathcal{H}_{A_1^i}\otimes\mathcal{H}_{A_2^i}$, and write
\begin{equation}\label{intrinsicguess}
\rho_{AB}= \bigoplus_i p_i\rho_{A_1^i}\otimes\rho_{A_2^iB}\,,
\end{equation}
where $p_i\ge0$, $\sum_ip_i=1$. We have $J(B,A)\le\sum_ip_iJ(B,A_2^i)-\sum_ip_i\ln p_i$, so if $J(B,A_2^i)$ is of order 1 for all $i$, and the mixing entropy $-\sum_ip_i\ln p_i$ is of order 1, then $J(B,A)$ is of order 1. Of course, the actual state $\rho_{AB}$ in a holographic theory may well take a form that is even more complicated than \eqref{intrinsicguess}.

\subsection{$S(B)+S(ABC)\le S(AB)+S(BC)$}\label{SSA1}

\emph{Proof:} Same strategy as for previous proofs, with $r'(B):=r(AB)\cap r(BC)$, $r'(ABC):=r(AB)\cup r(BC)$. $\Box$

This inequality is called \emph{strong subadditivity} (SSA). It says that the mutual information increases under inclusion, $I(A:BC)\ge I(A:B)$, as we would expect from a measure of correlation. By setting $B=\emptyset$, it implies subadditivity (\ref{subadditivity}). It is a general property of entropy in any quantum system, and is therefore required for consistency.

The proof above was essentially given in \cite{Headrick:2007km}, except that in the decomposition of the surfaces $m(AB)$, $m(BC)$, $\partial r'(AB)$, $\partial r'(BC)$, the terms $\sam(r(AB),r(BC))$, $\opp(r(AB),r(BC))$ were neglected. These terms can be finite, even without any fine-tuning of the geometry. For example, if $B$ has a component $B_1$ that is distant from $A$, $C$, and the other components of $B$, then $r(B_1)$ will be a component of both $r(AB)$ and $r(BC)$, so $\sam(r(AB),r(BC))$ will include $S(B_1)$. Similarly, if $A$ includes a small component $A_1$ that is surrounded by $C$, then $\opp(r(AB),r(BC))$ will include $S(A_1)$. Aside from allowing us to complete the proof of the SSA inequality, recognizing the presence of these terms is important when investigating the conditions for its saturation.

\subsubsection{Saturation}\label{SSA1saturation}

\emph{Condition:} $S(B)+S(ABC)=S(AB)+S(BC)$ if and only if $r(ABC)=r(AB)\cup r(BC)$, $r(B)=r(AB)\cap r(BC)$, $\opp(r(AB),r(BC))=0$.

\emph{Proof:} Follows directly from the proof of SSA. $\Box$

\begin{figure}[tbp]
\centering
\includegraphics[width=0.8\textwidth]{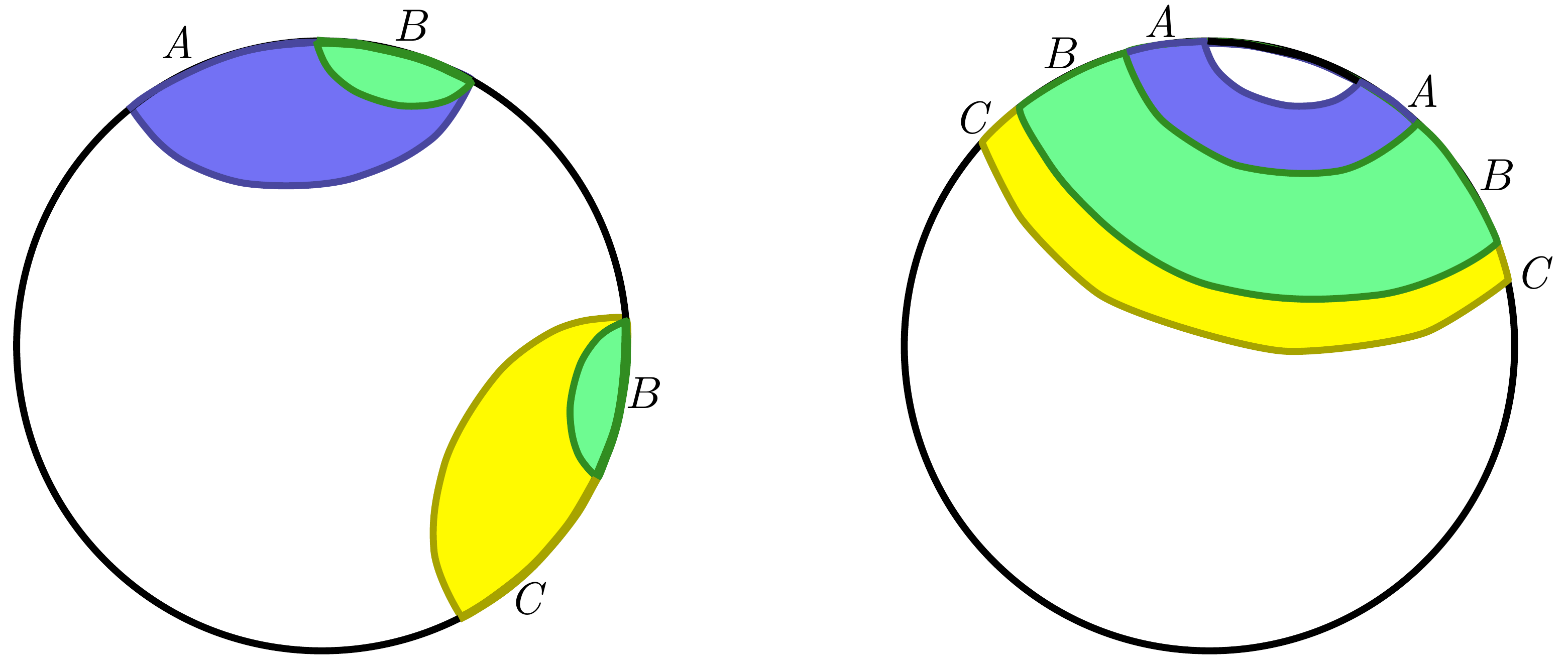}
\caption{\label{fig:SSAsaturation}
Illustration of cases where SSA is saturated. In each diagram, $r(B)$ is the green region, $r(AB)$ is blue + green, $r(BC)$ is green + yellow, and $r(ABC)$ is blue + green + yellow.}
\end{figure}

Figure \ref{fig:SSAsaturation} illustrates two kinds of configurations in which SSA can be saturated. (More complicated configurations are also possible, for example by combining these two.) The configuration on the left resembles the one in figure \ref{fig:subsaturation}: if we decompose $B$ as $B=B_1B_2$, where the subregions $B_{1,2}$ are close to $A$ and $C$ respectively, then we have $I(AB_1:B_2C)=0$. In the configuration on the right side, on the other hand, $B$ cannot be decomposed spatially into parts that are close to $A$ and $C$ respectively.

The bulk regions $r(A)$ and $r(C)$ are not shown in figure \ref{fig:SSAsaturation} (as their surface areas don't enter into the SSA inequality). However, according to Properties \ref{monotonicity} and \ref{nonoverlapping}, $r(A)$ must be contained in the blue region; similar with $r(C)$ and the yellow region. It follows that $I(A:C)=0$. In fact, this is always the case: using Properties \ref{subadditivity} and \ref{MMI}, it is easy to show that  $S(B)+S(ABC)=S(AB)+S(BC)$ implies $I(A:C)=0$.

As with subadditivity and AL, we should ask what form the reduced density matrix $\rho_{ABC}$ takes when SSA is saturated. Again, we begin by asking what its (exact) saturation implies for $\rho_{ABC}$ in a general quantum system, and, again, a theorem is available to answer that question. Saturation of SSA is equivalent to $I(A:BC)=I(A:B)$, so we would expect that $C$ has no correlations with $A$ other than ones that are already present in the $AB$ system. The following theorem \cite{HaydenJPW04} shows that this intuition is correct: $I(A:BC)=I(A:B)$ if and only if there exists a decomposition of the $B$ Hilbert space,
\begin{equation}
\mathcal{H}_B = \bigoplus_i\mathcal{H}_{B_1^i}\otimes\mathcal{H}_{B_2^i}
\end{equation}
such that
\begin{equation}\label{QMC}
\rho_{ABC} = \bigoplus_ip_i\rho_{AB_1^i}\otimes\rho_{B_2^iC}\,,
\end{equation}
where $p_i\ge0$, $\sum_ip_i=1$. In a state of this form, which is called a \emph{quantum Markov chain}, the correlations between $A$ and $C$ are entirely mediated by $B$.

Of course, just like subadditivity and AL, SSA is presumably not saturated exactly in holographic theories, but only at order $1/G_{\rm N}$ (except in trivial cases), so the form \eqref{QMC} should not be taken literally. Nonetheless, it is quite suggestive, and may be approximately true in some sense. In particular, it is clear that, in a configuration of the kind shown on the left side of figure \ref{fig:SSAsaturation}, the $B$ Hilbert space can be decomposed geometrically as $\mathcal{H}_B=\mathcal{H}_{B_1}\otimes\mathcal{H}_{B_2}$, with $\rho_{ABC}$ containing only an order-1 amount of classical correlation and entanglement between $AB_1$ and $B_2C$. On the other hand, in a configuration of the kind shown on the right side, any such decomposition cannot be (only) geometrical, must be (also) be somehow among the ``glueball'' degrees of freedom, just as in the saturation of AL.

\subsection{$S(A)+S(C)\le S(AB)+S(BC)$}\label{SSA2}

\emph{Proof:} Same strategy as for previous proofs, with $r'(A):=r(AB)\setminus r(BC)$, $r'(C):=r(BC)\setminus r(AB)$. $\Box$

This inequality is also a form of SSA. It says that the intrinsic entropy increases under inclusion (with the total system fixed), $J(AB,C)\ge J(A,BC)$, as we would expect. By setting $C=\emptyset$, it implies AL \ref{ArakiLieb}. It is a general property of entropy in any quantum system, and is therefore required for consistency.

\subsection{$S(A)+S(B)+S(C)+S(ABC)\le S(AB)+S(BC)+S(AC)$}\label{MMI}

{\emph Proof:} Same strategy as for previous proofs, with $r'(A):=r(AB)\cap r(AC)\setminus r(BC)$, etc. and $r'(ABC):=r(AB)\cup r(BC)\cup r(AC)$. $\Box$

This property is called \emph{monogamy of mutual information} (MMI). It can be written in terms of the mutual information as $I(A:BC)\ge I(A:B)+I(A:C)$ (in which form it resembles inequalities of a general class called \emph{monogamy inequalities}), or in terms of the intrinsic entropy as $J(AB,C)\ge J(A,BC)+J(B,AC)$.

The proof above was given in  \cite{Hayden:2011ag}, except that terms from coincident boundaries, such as $\opp(r(AB),r(BC))$, were neglected.

Unlike the previous four properties, MMI is not a general property of quantum systems;\footnote{However, the following similar but weaker inequality can be derived from SSA, and is therefore a general property of quantum systems: $S(A)+S(B)+S(C)+S(ABC)\le \frac43(S(AB)+S(BC)+S(AC))$.} for example, it is violated by the state on three bits $\rho_{ABC}=\frac12(\ket{000}\bra{000}+\ket{111}\bra{111})$. It is also violated in many quantum field theories \cite{CasiniHuerta2009}. Thus it is not required for consistency, but is rather a special property of holographic systems.\footnote{Intringuingly, it also seems to be obeyed by massive $2+1$ dimensional theories with long-range topological order \cite{Hayden:2011ag}.} (Even in holographic theories, it can likely be violated by order-1 corrections when it is saturated at order $1/G_{\rm N}$.) Its physical interpretation is not entirely clear, but it seems to indicate that the correlations between spatial regions in holographic theories are dominated by entanglement rather than classical correlation, and, perhaps more importantly, that this continues to be true even after tracing out other regions. In general, tracing out one part of a system decoheres the rest, converting entanglement into classical correlations. For example, if $ABC$ consists of three qubits in the entangled state $
\rho_{ABC}=\frac12(\ket{000}+\ket{111})(\bra{000}+\bra{111})$, then tracing out $C$ leads to the classically correlated state $\rho_{AB}=\frac12(\ket{00}\bra{00}+\ket{11}\bra{11})$. Apparently this does not happen in holographic theories, at least at leading order in $G_{\rm N}$. Perhaps the fact that such theories have a large number of degrees of freedom at each spatial point allows them to remain dominantly entangled despite the decoherence that occurs as a result of tracing out regions. A fuller discussion can be found in \cite{Hayden:2011ag}.

MMI, together with SSA, implies an infinite set of constrained inequalities on four or more subsystems that hold for any quantum system (but are independent of SSA) \cite{LindenWinter2005,Cadney2011}. Since these are required for consistency, they provide a remarkably stringent test of the RT formula. The simplest of them is as follows: If $I(A:BC)=I(A:B)=I(A:C)$ and $I(B:CD)=I(B:C)$, then $I(C:D) \geq I(C:AB)$.

\subsection{$S(A\bar A)+S(B\bar B)\le S(A\bar B)+S(B\bar A)$}\label{reflection}

We will call this property the \emph{reflection inequality}. Unlike the previous inequalities in this section, this is a new result. Also, the set-up and notation are slightly different than before, as we will explain.

Consider first a field theory on Minkowski space with coordinates $x^\mu$. Let $A,B$ be spacelike regions lying in the ``left'' Rindler wedge $x^1<0$, $|x^0|<|x^1|$. $A,B$ need not be disjoint, or even lie on a common spacelike slice. Let $\bar A,\bar B$ be the regions in the ``right'' Rindler wedge $x^1>0$, $|x^0|<|x^1|$ obtained by acting on $A,B$ respectively with a simultaneous time reversal and parity transformation,
\begin{equation}
(x^0,x^1,x^2,\ldots)\to(-x^0,-x^1,x^2,\ldots)\,.
\end{equation}
Casini has conjectured that the vacuum EEs in any unitary field theory obey the above reflection inequality \cite{Casini:2010nn}. The conjecture was motivated by an analogy between EEs and correlators, together with a property of correlators called wedge reflection positivity, which is a Lorentzian analogue of reflection positivity. Note that the reflection symmetry implies that $S(B\bar A)=S(A\bar B)$, so the inequality can also be written
\begin{equation}\label{RI2}
\frac12\left(S(A\bar A)+S(B\bar B)\right)\le S(A\bar B)\,.
\end{equation}
Or it can be written in terms of the mutual informations:
\begin{equation}\label{reflection2}
\frac12\left(I(A:\bar A)+I(B:\bar B)\right)\ge I(A:\bar B)\,.
\end{equation}

We will give a sufficient condition, in terms of the bulk geometry, for a state in a holographic theory to obey the reflection inequality. The condition is that $\Sigma$ (and hence $\dot\Sigma)$ admits a reflection symmetry, that is, a $\Z_2$ isometry with a fundamental domain $W$ such that $w:=\partial W\cap\partial\bar W$ is the fixed locus of the isometry (where the bar indicates the action of the isometry).\footnote{More generally, this reflection can be thought of the restriction to $\Sigma$ of a combined CPT transformation on the full spacetime $M$ and all the fields in it. Invariance of $M$ under this CPT transformation is equivalent to the state $\rho$ of the field theory being CPT-invariant. However, only the parity invariance of the metric on $\Sigma$ will play a role in our considerations.} Under this assumption, we will show that \eqref{RI2} holds for $A,B\subseteq\dot W$. (Again, unlike in the rest of the paper, $A,B$ need not be disjoint. We will never consider $S(AB)$, but only $S(A\bar A),S(B\bar A),S(A\bar B)$.) In particular, the condition is obeyed by the vacuum of a holographic theory on Minkowski space, since $\Sigma$ in that case is hyperbolic space, which, in the usual Poincar\'e coordinates, is invariant under $x^1\to-x^1$. Hence our theorem includes, as a special case, Casini's conjecture applied to a constant-time slice of a holographic theory, and thereby supports its validity. Conversely, if Casini's conjecture is generally true, then this property (applied to the vacuum of a theory on Minkowski space) is required for consistency of the RT formula.

Before giving the proof, we would first like to show that the reflection inequality is independent of the previous ones in this section, by giving an example of a state that violates it but obeys the others. Let $A,B$ be disjoint, and let
\begin{equation}
\rho_{AB\bar A\bar B} = \rho_{A\bar B}\otimes\rho_{B\bar A}\,,
\end{equation}
where $\rho_{A\bar B}$ and $\rho_{B\bar A}$ do not factorize.\footnote{It is not clear if such a state can exist in a holographic theory with a connected bulk $\Sigma$. If it can, then of course by our theorem the bulk would not admit a reflection symmetry.} It is straightforward to check that this state obeys (in fact saturates) MMI; being an allowed quantum state it necessarily satisfies all of the other inequalities of this section as well. However, it violates the reflection inequality, as the left-hand side of \eqref{reflection2} vanishes while the right-hand side is positive.

\emph{Proof:} We assume that $r(A\bar B),r(B\bar A)$ are related by a reflection (if the minimizers are unique, then this must be the case; if not, choose them so). We define the regions
\begin{equation}
r_1:= r(A\bar B)\cap W\,,\qquad r_2:=r(B\bar A)\cap W\,,
\end{equation}
so that we have $r(A\bar B)=r_1\cup\bar r_2$. From \eqref{unionlemma}, and noting that $r_1$ and $\bar r_2$ are disjoint, we have
\begin{equation}
S(A\bar B)=\ext(r_1,\bar r_2) + \ext(\bar r_2,r_1)\,.
\end{equation}
We can further decompose $\ext(r_1,\bar r_2)$ into the part of $\partial r_1$ exterior to $\bar W$ and the part along $w$:
\begin{equation}
\ext(r_1,\bar r_2) = \ext(r_1,\bar W) + \area(\partial r_1\cap w\setminus\partial\bar r_2)\,;
\end{equation}
similarly with $\ext(\bar r_2,r_1)$. All in all, and using the symmetry, we find
\begin{equation}
S(A\bar B) =
\ext(r_1,\bar W)+\ext(r_2,\bar W)+\area(\partial r_1\cap w\setminus\partial\bar r_2)+\area(\partial r_2\cap w\setminus\partial\bar r_1)\,.
\end{equation}

\begin{figure}[tbp]
\centering
\includegraphics[width=0.7\textwidth]{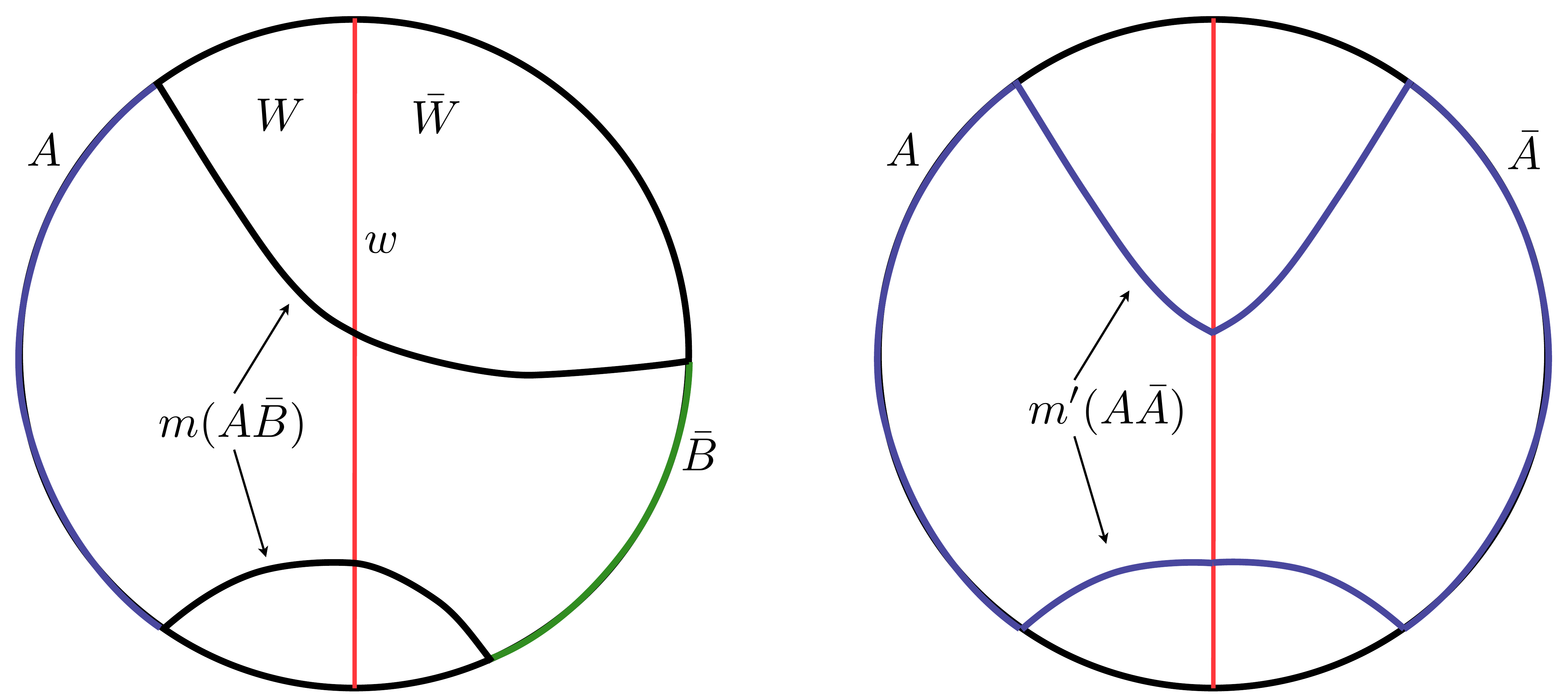}
\caption{\label{fig:reflection}
Left: Reflection-symmetric bulk $\Sigma$, divided into two fundamental domains $W,\bar W$ separated by the fixed locus $w$ of the reflection. Boundary regions $A\subseteq\dot W$ and $\bar B\subseteq\dot{\bar W}$ are indicated, along with the bulk surface $m(A\bar B)$. Right: The surface $m'(A\bar A):=\partial r'(A\bar A)$ used in the proof of Property \ref{reflection}, for the surfaces shown on the left.}
\end{figure}

We now define the regions
\begin{equation}
r'(A\bar A):=r_1\cup\bar r_1\,,\qquad r'(B\bar B):=r_2\cup\bar r_2
\end{equation}
and their boundary areas $S'(A\bar A)$, $S'(B\bar B)$ (see figure \ref{fig:reflection}). We have a similar decomposition as for $S(A\bar B)$. However, the terms $\area(\partial r_1\cap w\setminus\partial\bar r_1)$ and $\area(\partial r_2\cap w\setminus\partial\bar r_2)$ vanish because $w$ is the fixed locus of the reflection. Hence we have
\begin{equation}
\frac12\left(S'(A\bar A)+S'(B\bar B)\right) = \ext(r_1,\bar W) +\ext(r_2,\bar W)\le S(A\bar B)\,.
\end{equation}
On the other hand, since $\dot r'(A\bar A)=A\bar A$ and $\dot r'(B\bar B)=B\bar B$,
\begin{equation}
S(A\bar A)\le S'(A\bar A)\,,\qquad S(B\bar B)\le S'(B\bar B)\,.
\end{equation}
$\Box$

\acknowledgments

I would like to thank A. Baskaran, P. Hayden, V. Hubeny, A. Lawrence, D. Marolf, and D. Ruberman for helpful conversations.  My work is supported in part by the National Science Foundation under CAREER Grant No.\ PHY10-53842.

\bibliographystyle{JHEP}
\bibliography{refs}

\end{document}